\documentclass[pra,aps,amsmath,showpacs,preprintnumbers]{revtex4}
\usepackage{bm}
\usepackage{graphicx}

\newcommand{\pt}{\partial}
\newcommand{\tc}{\textcircled}
\newcommand{\mL}{\mathcal{L}}
\newcommand{\mT}{\mathcal{T}}
\newcommand{\mE}{\mathcal{E}}

\begin{document}

\title{Dynamics of a single ring of vortices
       in two-dimensional trapped Bose-Einstein condensates}

\author{Jong-kwan Kim$^{1,3}$ and Alexander L. Fetter$^{1,2,3}$ }
\affiliation{$^{1}$Geballe Laboratory for Advanced Materials,
                   Stanford University,
                   Stanford, CA 94305-4045\\
             $^{2}$Department of Physics,
                   Stanford University,
                   Stanford, CA 94305-4060\\
             $^{3}$Department of Applied Physics,
                   Stanford University,
                   Stanford, CA 94305-4090 }

\date{\today}

\begin{abstract}
The dynamics of a ring of vortices in two-dimensional Bose-Einstein condensates
(with and without an additional vortex at the center) is studied
for (1) a uniform condensate in a rigid cylinder
and (2) a nonuniform trapped condensate in the Thomas-Fermi limit.
The sequence of ground states (within these single-ring configurations)
is determined as a function of the external rotation frequency
by comparing the free energy of  the various states.
For each ground state, the Tkachenko-like excitations
and the associated dynamical stability are analyzed.
\end{abstract}


\maketitle


\section{introduction}
  The dynamics of quantized vortices has attracted great interest.
Until recently most research has been done
in the context of superfluid $^4$He~\cite{Donn91}.
The physics of an unbounded superfluid
with an infinite number of vortices is relatively simple:
In equilibrium the vortices form
a two-dimensional rotating triangular vortex lattice,
which supports small-amplitude collective modes~\cite{Tkac66}.
In a bounded superfluid, however,
determining the equilibrium configuration is not easy
even for a small number of vortices.

  Havelock~\cite{Have31} studied the dynamical stability
of a ring with a small number of vortices.
Subsequently, Hess~\cite{Hess67} studied
the energy of various configurations analytically
to determine the sequence of low-lying equilibrium states,
and Campbell and Ziff~\cite{Camp79} evaluated the energies
numerically for larger numbers of vortices.
After heroic efforts, Yarmchuk {\it et al}.~\cite{Pack79}
photographed some stable vortex patterns in superfluid $^4$He.

  For a large number of vortices,
Fetter developed a theory of the dynamics
of a lattice of slightly displaced rectilinear vortices~\cite{Fett67}
including three-dimensional distortions.
Later, the description was extended to provide a continuum picture
of the vortex dynamics~\cite{Fett77}.
Baym and Chandler formulated similar macroscopic hydrodynamics
of rotating superfluids~\cite{Baym83}.
These various theoretical descriptions included
the vortex-interaction effects discussed by Tkachenko~\cite{Tkac66}.
For a comprehensive review of vortex dynamics in rotating superfluids,
see, for example, Ref.~\cite{Soni87}.

  The strong interaction between helium atoms makes it difficult
to manipulate the condensate and to observe vortices.
Thus, the detection of quantized vortices
in trapped Bose-Einstein condensates~\cite{Matt99,Dali00,Madi00}
offers an opportunity to study the dynamics of superfluid vortices.
In addition, the interaction parameter is tunable 
by varying the number of atoms, 
which helps clarify their many-body character
and the detailed connection to fluid dynamics.
Recent experiments at JILA~\cite{Codd03}
measured the transverse elastic modes of the bounded vortex lattice
that have been studied theoretically by \cite{Angl02,Baym03,Stri04,Soni04}.
These Tkachenko modes~\cite{Tkac66}
in a large but finite lattice
require appropriate and nontrivial boundary conditions.

  Here, we study the equilibrium states and small oscillations
of (1) a ring of $N_v$ vortices
and (2) a ring of $N_v-1$ vortices plus a central vortex
for two specific cases:
a cylindrical container with uniform fluid
and a two-dimensional condensate in a harmonic radial trap
in the Thomas-Fermi (TF) limit.
These vortex states can be stable up to $N_v\sim9$,
beyond which other more complicated configurations become favorable.
The associated normal modes are small-system analogs
of the Tkachenko modes for a large vortex array.

\section{time-dependent variational formalism}
  A macroscopic order parameter (the condensate wave function)
$\Psi\equiv\langle\hat{\Psi}\rangle$,
with associated particle density $n=\left|\Psi\right|^2$,
provides a simple description of a Bose-Einstein condensate near $T=0$ K.
In the presence of an external trap potential $V_{\mathrm{tr}}$,
the condensate experiences both $V_{\mathrm{tr}}$
and the self-consistent Hartree interaction $V_{\mathrm{H}}=g|\Psi|^2=gn$
arising from the interaction with all other particles,
where $g=4\pi\hbar^2 a_{s}/M$ is
the interparticle interaction parameter (assumed to be repulsive)
and $a_{s}$ is the positive $s$-wave scattering length~\cite{Dalf99,Fett99}.

  At zero temperature in the laboratory (nonrotating) frame, 
the macroscopic order parameter $\Psi$ satisfies
the time-dependent, nonlinear Schr\"odinger equation
(Gross-Pitaevskii equation),
\begin{equation}
\label{eq:GP}
i\hbar\frac{\pt\Psi}{\pt t}
   = - \frac{\hbar^2\nabla^2}{2M}\Psi
     + V_{\mathrm{tr}}\Psi
     + g \left|\Psi\right|^2 \Psi.
\end{equation}
If the external potential rotates at an angular velocity $\bm{\Omega}$, 
it is necessary to transform to a co-rotating frame, 
in which case the time-dependent Schr\"odinger equation becomes~\cite{Landau} 
\begin{equation}
\label{eq:GP-rot}
i\hbar\frac{\pt\Psi}{\pt t}
   = - \frac{\hbar^2\nabla^2}{2M}\Psi
     + V_{\mathrm{tr}}\Psi - \bm{\Omega}\cdot\mathbf{r}\times\mathbf{p}\,\Psi
     + g \left|\Psi\right|^2 \Psi,
\end{equation}
where $\mathbf{r}\times\mathbf{p}= -i\hbar \mathbf{r}\times\bm{\nabla}$ 
is the angular-momentum operator.

  For many purposes, however,
it is more convenient to use an equivalent Lagrangian formalism
based on the Lagrangian functional~\cite{Pere96},
\begin{equation}
\mL[\Psi] \equiv \mT[\Psi] - \mE'[\Psi].
\end{equation}
Here
\begin{equation}
\mT[\Psi]
   = \int dV\, \frac{i\hbar}{2}
     \left(\Psi^*\frac{\pt\Psi}{\pt t}
           - \frac{\pt\Psi^*}{\pt t}\Psi
     \right),
\end{equation}
is the time-dependent part of the Lagrangian,
the analog of the kinetic energy in classical mechanics.
In a nonrotating frame, $\mE[\Psi]$ is the energy functional.
In the reference frame that rotates at the same angular velocity
$\mathbf{\Omega}$ as the external trap potential,
the appropriate free energy functional acquires an additional term,
$\mE'[\Psi] = \mE[\Psi] - \mathbf{\Omega}\cdot\mathbf{L}[\Psi]$,
where $\mathbf{L}[\Psi]$ is the angular momentum of the condensate.
When $\bm{\Omega}=\Omega\hat{\mathbf{z}}$ is oriented along
$\hat{\mathbf{z}}$,
$\mathcal{E}'$ can be written as
\begin{equation}
\mE'[\Psi]
   = \int dV
     \left(\frac{\hbar^2}{2M}\left|\bm{\nabla}\Psi\right|^2
           + V_{\mathrm{tr}} \left|\Psi\right|^2
           + \frac{g}{2} \left|\Psi\right|^4
           + i\hbar\Omega \Psi^* \frac{\pt \Psi}{\pt \phi}
     \right).
\end{equation}

  The action $\mathcal{S}$ associated with the above Lagrangian
is the time integral $\mathcal{S}[\Psi]=\int dt~\mL[\Psi]$.
It is stationary with respect to small variations of $\Psi$ and $\Psi^*$,
when $\Psi$ satisfies
the time-dependent Gross-Pitaevskii (GP) equation~(\ref{eq:GP-rot}).
Thus the time-dependent GP equation is the Euler-Lagrange equation
for the Lagrangian functional.

  This Lagrangian formalism has the following significant advantage:
not only is this approach exact,
but also it provides the basis for a powerful approximate variational method.
If the condensate wave function depends
on various parameters $\{\mathbf{p}\}$ of the given system,
the resulting Lagrangian $\mL(\{\mathbf{p}\})$ provides
the dynamical equations of motion of these parameters $\{\mathbf{p}\}$.
Moreover, when the system under consideration is in equilibrium
and the variational parameters of the problem are stationary,
the next-order perturbation (variation) of the Lagrangian yields
the normal modes of the system in terms of these parameters.

  Here we use this method to study the dynamics of vortices in a ring
in a two-dimensional Bose condensate.
The positions of the vortices \{$\mathbf{r}_j$\} serve
as the time-dependent variational parameters
for two distinct cases:
(1) a uniform condensate in a rigid cylinder
and (2) a nonuniform trapped condensate in the TF limit~\cite{Baym96}.

  In our variational approach,
we use a trial wave function of the following form
\begin{equation}
\label{eq:trial}
\Psi(\mathbf{r},\{\mathbf{r}_j\})
   = \left|\Psi(\mathbf{r})\right| e^{i S(\mathbf{r},\{\mathbf{r}_j\})},
\end{equation}
where $\left|\Psi\right|^2$ is the density profile of
either the uniform condensate or the nonuniform TF condensate.
It satisfies the normalization condition
\begin{equation}
N=\int dV \left|\Psi\right|^2,
\end{equation}
where $N$ is the number of condensate particles.
In addition, $S(\mathbf{r},\{\mathbf{r}_j\})$
is the phase of the condensate wave function,
and the particle velocity field is given by
$\mathbf{v}(\mathbf{r})=\left(\hbar/M\right)\bm{\nabla} S(\mathbf{r})$.
The trial wave function~(\ref{eq:trial}) assumes
that the vortices do not affect the density profile of the condensate
(this approximation holds for small cores, which is valid in the TF limit);
instead, the vortices directly determine the velocity field.
For simplicity, we consider only a condensate in two dimensions,
when the various terms in the Lagrangian
are interpreted per unit length.

\section{free energy and equilibrium states}
  To proceed, it is first necessary to choose a specific form
for the phase $S$ of the condensate wave function.
The boundary conditions are quite different
for a uniform condensate and for a TF condensate,
and it is convenient to treat them separately.

\subsection{Uniform condensate in a cylinder}
  The uniform condensate in a rigid circular cylinder of radius $R$
must satisfy the boundary condition
that the normal component of the particle current
vanishes at the surface of the cylinder.
This condition can be ensured with an image vortex
of opposite circulation at the external position
$\mathbf{r}_j'=\left(R^2/r^2_j\right)\mathbf{r}_j$
for each vortex at $\mathbf{r}_j$ in the cylinder.

  The time-dependent part $\mT$ of the Lagrangian can be easily evaluated.
The trial condensate wave function is
\begin{equation}
\Psi(\mathbf{r},\{\mathbf{r}_j\})
   = \sqrt{n} e^{i S(\mathbf{r},\{\mathbf{r}_j\})},
\end{equation}
and
\begin{equation}
S(\mathbf{r},\{\mathbf{r}_j\})
   = \sum_{j=1}^{N_v} S_j(\mathbf{r},\mathbf{r}_j)
     + \sum_{j=1}^{N_v} S_j'(\mathbf{r},\mathbf{r}_j'),
\end{equation}
where
\begin{align}
S_j(\mathbf{r},\mathbf{r}_j)
   &= \arctan\left(\frac{y-y_j}{x-x_j}\right), \\
S_j'(\mathbf{r},\mathbf{r}_j')
   &= (-1) \arctan\left(\frac{y-y_j'}{x-x_j'}\right),
\end{align}
$n=N/V$ is the particle density
and $N_v$ is the number of the vortices in the condensate.
In cylindrical coordinates, we obtain
\begin{equation}
\label{eq:Tunif}
\mT(\{\mathbf{r}_j\})
   = \sum_{j=1}^{N_v} \mT_j(\mathbf{r}_j)
   = \sum_{j=1}^{N_v} \left(1-r_j^2\right) \dot{\phi}_j,
\end{equation}
where each quantity is per particle and written in dimensionless units:
the length scaled by the radius of the cylinder $R$,
the time scaled by $\Omega_0^{-1}\equiv MR^2/\hbar$
and the energy by $\hbar\Omega_0$.

  In the free energy part of the Lagrangian,
the quantity of interest is
the {\it difference} $\Delta\mE'$ of the free energy
between the vortex state and the vortex-free state.
In our approximation, this difference arises
from the superfluid velocity field $\mathbf{v}$
caused by the presence of the vortices.
It can be simply determined as
\begin{equation}
\Delta\mE'=\Delta\mE-\Omega L_z,
\end{equation}
where $\Delta\mE$ is the contribution from the superfluid velocity field
in the nonrotating reference frame
\begin{equation}
\label{eq:kinetic}
\Delta\mE=\int d^2\mathbf{r} \left|\Psi\right|^2 \frac{1}{2}\,\mathbf{v}^2,
\end{equation}
and $L_z$ is the $z$ component of the angular momentum
\begin{equation}
L_z = \int d^2\mathbf{r}
        \left|\Psi\right|^2 \hat{\mathbf{z}}\cdot\mathbf{r}\times\mathbf{v}.
\end{equation}

  In evaluating the free energy part,
it is convenient to use a stream function $\chi$
instead of the phase of the condensate wave function $S$
(which is effectively a velocity potential).
Thus, we choose
\begin{equation}
\chi(\mathbf{r},\{\mathbf{r}_j\})
  = \sum_{j=1}^{N_v} \chi_j(\mathbf{r},\mathbf{r}_j)
    + \sum_{j=1}^{N_v} \chi_j'(\mathbf{r},\mathbf{r}_j'),
\end{equation}
where
\begin{align}
\chi_j(\mathbf{r},\mathbf{r}_j)
   &= \ln\left|\mathbf{r}-\mathbf{r}_j\right|, \\
\chi'_j(\mathbf{r},\mathbf{r}_j')
   &= (-1) \ln\left|\mathbf{r}-\mathbf{r}_j'\right|,
\end{align}
and the velocity field is given by
\begin{equation}
\label{eq:stream}
\mathbf{v}(\mathbf{r})=\hat{\mathbf z}\times\bm{\nabla}\chi.
\end{equation}
This two-dimensional stream function has the advantage
being single-valued.
When describing irrotational flow in the presence of vortices,
$\chi$ satisfies Poisson's equation
\begin{equation}
\label{eq:Poisson}
\nabla^2\chi(\mathbf{r},\{\mathbf{r}_j\})
  = 2\pi \sum_{j=1}^{N_v} \delta^{(2)}\left(\mathbf{r}-\mathbf{r}_j\right).
\end{equation}
The right-hand side vanishes everywhere
except for the finite number of singular points
at the location of the vortices 
(the image vortices lie outside the physical region and  do not appear here).
Thus, this stream function is essentially
a superposition of two-dimensional Green's functions.

  Each individual term $\chi_j(\mathbf{r},\mathbf{r}_j)$ 
has the Fourier expansion,
\begin{equation}
\label{eq:harmonics}
\chi_j(\mathbf{r},\mathbf{r}_j)
   = \Theta(r-r_j)
       \left[\ln r
             - \sum_{m=1}^\infty
                 \frac{\cos m\phi}{m} \left(\frac{r_j}{r}\right)^m
       \right]
   + \Theta(r_j-r)
       \left[\ln r_j
             - \sum_{m=1}^\infty
                 \frac{\cos m\phi}{m} \left(\frac{r}{r_j}\right)^m
       \right],
\end{equation}
where
\begin{equation}
\Theta(r-r_j) = \begin{cases}
                     1, \qquad \text{if}~~ r>r_j, \\
                     0, \qquad \text{if}~~ r<r_j,
                 \end{cases}
\end{equation}
is the unit positive step function
and $\phi$ is the angle between $\mathbf{r}$ and $\mathbf{r}_j$.
Straightforward analysis then yields the free energy difference~\cite{Hess67},
\begin{equation}
\label{eq:freeE_unif}
\begin{split}
\Delta \mE'(\{\mathbf{r}_j\})
   = & N_v\ln\left(\frac{R}{\xi}\right)
              + \sum_{j=1}^{N_v} \ln\left(1-r_j^2\right)
              - \Omega \sum_{j=1}^{N_v} \left(1-r_j^2\right) \\
     &\phantom{~}
              + \sum_{i<j}^{N_v} \ln\left(\frac{1-2r_i r_j
\cos\phi_{ij}+r_i^2 r_j^2}
                                               {r_i^2-2r_i r_j
\cos\phi_{ij}+r_j^2}
                                    \right),
\end{split}
\end{equation}
where $\xi$ is the healing length of the condensate
(the vortex-core-radius cutoff)
and $\phi_{ij}\equiv\phi_i-\phi_j$ is the angle
between the $i$th and the $j$th vortices.
Each vortex interacts with all other vortices including images,
as is evident from the last term.
Note that $\Omega$ is also written in dimensionless units scaled with $\Omega_0$.

\subsection{TF condensate}
  In the TF limit, 
it is more convenient to work with the equivalent grand canonical ensemble,
introducing the chemical potential $\mu$
that is determined by the normalization condition
$N=\int dV \left|\Psi\right|^2$.
Then the condensate wave function gives the following TF density profile
\begin{equation}
\left|\Psi(\mathbf{r})\right|^2 = n_0 \left(1-r^2\right),
\end{equation}
where $n_0=\mu/g$ is the central density
and $r$ is a dimensionless radius,
scaled by the TF radius $R=\sqrt{2\mu/M\omega_{\perp}^2}$
(in the TF limit, $\mu \gg \hbar\omega_\perp$,
which implies that $R$ is large compared
to the oscillator length $\sqrt{\hbar/M\omega_\perp}$).

  In the TF case, the condensate density vanishes at the boundary ($r=1$).
Thus the particle current automatically satisfies the boundary condition,
and image vortices are unnecessary.
When the energy is scaled by the characteristic energy
$\hbar\Omega_0=\hbar^2/MR^2=(\hbar\omega_\perp/2\mu)\,\hbar\omega_\perp$,
the time-dependent part of the Lagrangian per particle is 
\begin{equation}
\label{eq:Ttf}
\mT(\{\mathbf{r}_j\})
   = \sum_{j=1}^{N_v} \mT_j(\mathbf{r}_j)
   = - \sum_{j=1}^{N_v} \left(r_j^2-\frac{r_j^4}{2}\right) \dot{\phi}_j.
\end{equation}

  The free energy difference per particle can be evaluated
using the same Fourier expansion~(\ref{eq:harmonics})
for the stream functions (see Appendix~\ref{ap:TFfree}),
which yields
\begin{equation}
\label{eq:freeE_TF}
\begin{split}
\Delta\mE'(\{\mathbf{r}_j\})
   &= \frac{1}{2}\sum_{j=1}^{N_v} \left\{\left(1-r_j^2\right)
                                     \left[2\ln\left(\frac{R}{\xi}\right)
                                           + \ln\left(1-r_j^2\right)\right]
                                     + \left(2r_j^2-1\right)
                                     - \Omega \left(1-r_j^2\right)^2 \right\} \\
   &\phantom{=~~}
     + \frac{1}{2}
       \sum_{i<j}^{N_v} \left(1-r_i r_j\cos\phi_{ij}\right)
                 \ln\left[\frac{1- 2r_i r_j \cos\phi_{ij} + r_i^2 r_j^2}
                               {\left(r_i^2 - 2r_i r_j \cos\phi_{ij} 
                                            + r_j^2\right)^2}
                    \right] \\
   &\phantom{=~~}
     + \sum_{i<j}^{N_v} r_i r_j\sin\phi_{ij}
                 \arctan\left(\frac{-r_i r_j\sin\phi_{ij}}
                                   {1- r_i r_j \cos\phi_{ij}}\right) \\
   &\phantom{=~~}
     - \sum_{i<j}^{N_v} \left(1-r_i^2-r_j^2\right).
\end{split}
\end{equation}

\subsection{Euler-Lagrange equations of motion}
\label{sec:Euler-Lagrange}
  The Euler-Lagrange equations for the Lagrangian $\mL=\mT-\mE'$ are
\begin{align}
\label{eq:Euler-Lagrange1}
0&=\frac{d}{dt}\frac{\pt\mL}{\pt\dot{r}_j}
    - \frac{\pt\mL}{\pt r_j}, \\
\label{eq:Euler-Lagrange2}
0&=\frac{d}{dt}\frac{\pt\mL}{\pt\dot{\phi}_j}
    - \frac{\pt\mL}{\pt \phi_j}.
\end{align}
Since $\mL$ is independent of $\dot{r}_j$ 
for the configurations studied here, 
the first equation reduces to $\pt\mL/\pt r_j = 0$, 
which determines the precession frequency of the $j$th vortex
[see Eqs.~(\ref{eq:Tunif}) and (\ref{eq:Ttf})],
\begin{equation}
\label{eq:prec_freq}
\dot{\phi}_j
   = \left(\frac{\pt\Delta\mE'}{\pt r_j}\right)
     \Bigg/
     \left(\frac{1}{\dot{\phi}_j}\,\frac{\pt\mT}{\pt r_j} \right).
\end{equation}
If the system is nondissipative
and all vortices lie symmetrically spaced on a single ring of radius $r$,
they are locked together and precess at the same rate
$\dot{\phi}=\dot{\phi}_j(r,\Omega)$.

\subsection{Uniform precession of a single vortex}
  It is interesting to start with the simplest case of a single vortex,
which clarifies the distinct role of the factor $\ln\left(R/\xi\right)$
between the Lagrangians for the uniform condensate and the TF condensate.
Consider the uniform condensate, 
where the free energy part of the Lagrangian~(\ref{eq:freeE_unif}) 
contains the logarithm $\ln\left(R/\xi\right)$
only as an {\it additive} constant $N_v \ln\left(R/\xi\right)$,
independent of the coordinates of the vortices.
Thus, this logarithmic factor $\ln\left(R/\xi\right)$ 
affects the energy-related properties,
such as the ground-state configuration
and the angular velocity for transitions between adjacent states;
in contrast, the same factor does not affect
either the radius of the ring of vortices or the precession frequency.
For example, Eq.~(\ref{eq:prec_freq}) gives
the scaled precession frequency for a single vortex at a distance $r_1$
in a uniform nonrotating condensate
\begin{equation}
\label{eq:prec_unif}
\dot{\phi}_1 = \frac{1}{1-r_1^2}.
\end{equation}
From a physical viewpoint,
the single vortex in a uniform condensate precesses 
under the influence of its opposite image vortex,
which explains why the precession rate increases
as the vortex moves toward the outer boundary.

  An expansion of $\Delta\mE'(r_1)$ from Eq.~(\ref{eq:freeE_unif})
for small $r_1$ near the center of the condensate shows
that the curvature changes sign at $\Omega_m = 1$.
This value is the same as the precession frequency
for a nearly central vortex in a nonrotating condensate;
$\Omega_m$ corresponds to the onset of metastability
for a central vortex~\cite{Pack72}.

  The situation is quite different for the TF condensate,
where Eq.~(\ref{eq:freeE_TF}) has $\ln\left(R/\xi\right)$
in a term that depends on the coordinates of the vortices.
Use of Eq.~(\ref{eq:prec_freq}) readily yields
the corresponding TF precession frequency for a single vortex
at $r_1$ in a nonrotating condensate
\begin{equation}
\label{eq:prec_TF}
\dot{\phi}_1
   = \frac{1}{1-r_1^2}
     \left[\ln\left(\frac{R}{\xi}\right)
           +\frac{1}{2}\ln(1-r_1^2)
           -\frac{1}{2}\right].
\end{equation}
Equation~(\ref{eq:prec_TF}) differs from Eq.~(\ref{eq:prec_unif})
because of the ``large'' factor $\ln\left(R/\xi\right)$,
which arises from the radial derivative of the energy
in the numerator of Eq.~(\ref{eq:prec_freq}).
Although both (\ref{eq:prec_unif}) and (\ref{eq:prec_TF})
have similar denominators,
the present factor $\left(1-r_1^2\right)^{-1}$ reflects 
the nonuniform TF density~\cite{Jack99,Lund00,McGee01,Fett01}
(instead of the image, since the TF condensate has no image vortices).
Physically, the precession here arises
from the combination of the Magnus force
and the gradient in the density
(or equivalently the trap potential 
in this TF limit)~\cite{Svid00,Svid00a,Shee04}.

  An expansion of $\Delta\mE'(r_1)$ from Eq.~(\ref{eq:freeE_TF})
for small $r_1$ near the center of the condensate shows
that the curvature now changes sign at
$\Omega_m = \ln(R/\xi)-\frac{1}{2}$,
which is again the same as the precession frequency 
for a nearly central vortex in a nonrotating condensate.
This value also corresponds to the onset of metastability,
now enhanced by the large logarithmic factor.

\subsection{State of lowest free energy}
  In the presence of any small dissipative process,
the system seeks the state of lowest free energy
for a given external rotation
by adjusting the radius of the ring of vortices to be $r_r$,
determined by the condition
\begin{equation}
\label{eq:r_r}
0 = \left. \frac{\pt\Delta\mE'}{\pt r_j}\right|_{\{r_j\}=r_r}.
\end{equation}
In this case, Eq.~(\ref{eq:prec_freq}) shows that $\dot\phi=0$,
which means that the position of each vortex does not change.
Hence all vortices are stationary in the co-rotating reference frame.
In the laboratory (inertial) frame,
the vortices all rotate at the same rate
as the external rotation $\Omega$.
Physically this result can be understood
when we imagine that the vortex core contains some normal component,
and the relative motion of the normal component
with respect to the frame of the rotating trap gives rise to dissipation.

  This equilibrium radius $r_r$ is
a function of the external rotation $\Omega$.
For a given vortex configuration,
we can evaluate the lowest free energy
$\Delta\mE'_{\mathrm{min}}(r_r,\Omega)$
by solving Eq.~(\ref{eq:r_r}).
Comparing these lowest free energies among the different vortex states,
we can determine the state with lowest energy
and the associated angular velocity $\Omega_c^{AB}$,
at which the transition from vortex state $A$
to vortex state $B$ should occur.
Here we will consider two types of vortex configurations
for each (uniform/TF) condensate;
a ring of vortices {\it without} a vortex at the center
(`$N_r$' configuration)
and a ring of vortices {\it with} a vortex at the center
(`$N_r$+\tc{1}' configuration).

  For both the uniform and the TF condensates, 
the vortex-free state is energetically favored for any angular velocity 
below $\Omega_c^{01}$ because it has the lowest free energy.
As $\Omega$ increases, however, 
the free energy of the one-vortex state decreases at a faster rate
than that of the vortex-free state.
At $\Omega_c^{01}$, the free energy of the two states is the same,
and the condensate can support a central vortex 
throughout the interval $\Omega_c^{01} \le \Omega \le \Omega_c^{12}$.
At higher $\Omega$ between $\Omega_c^{12}$ and $\Omega_c^{23}$,
the state with two vortices has the lowest energy, and so on.

  The two condensates display one interesting difference, however, 
arising from the different density profiles.
In the TF approximation, 
the thermodynamic critical angular velocity $\Omega_c^{01}$ 
(the usual lower critical angular velocity $\Omega_{c1}$)
for the appearance of the first central vortex is given very generally by
\begin{equation}
\Omega_c^{01} 
  = \frac{\Delta\mE_1}{L_1} 
  = \frac{\int d^2 r\,\frac{1}{2} v^2 n(r)}
         {\int d^2 r \,n(r)} 
  = \frac{1}{2} 
    \frac{\int d^2 r\,r^{-2} n(r)}{\int d^2 r\,n(r)},
\end{equation}
where $\Delta\mE_1$ is the energy per particle for one central vortex 
and $L_1$ is the angular momentum per particle for the same central vortex.
Evidently, $\Omega_c^{01}$ is proportional to the average value of $r^{-2}$ 
with the number density as the non-negative weight. 
The TF condensate $n_{TF}(r) \propto 1-r^2$ 
has the maximum density at the center; 
thus the corresponding $\Omega_c^{01}$ is larger 
than that for the uniform condensate
(the difference leads to a factor $\approx 2$ 
in the present two-dimensional case).

  The state with five vortices in a ring plus one at the center
(5+\tc{1} configuration) always has a lower free energy
than the state with all six vortices in a ring (6 configuration),
so the `5' to `6' transition is not favored; instead, `5' goes to `5+\tc{1}'.
In fact, this specific `5' to `5+\tc{1}' transition was observed 
in an early experiment on rotating superfluid $^4$He~\cite{Pack79}.
In a uniform unbounded condensate,
the vortices form a triangular (Abrikosov) lattice
with six-fold rotational symmetry.
Here the transition from a single ring with an empty center (`$N_r$')
to a ring with a central vortex (`$N_r$+\tc{1}') happens
when the condensate has five vortices in a ring ($N_r=5$) instead of six,
which reflects the finite size of our geometry.

\begin{figure}[h]
\centering
\begin{tabular}{cc}
\begin{minipage}{3in}
\includegraphics[height=2.4in,width=3in]{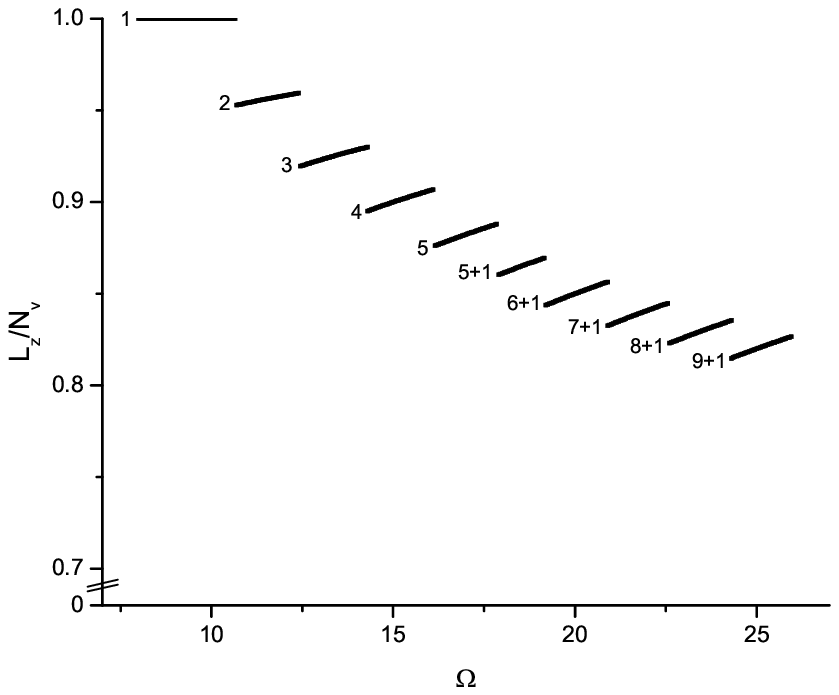}
\end{minipage}
&
\begin{minipage}{3in}
\includegraphics[height=2.4in,width=3in]{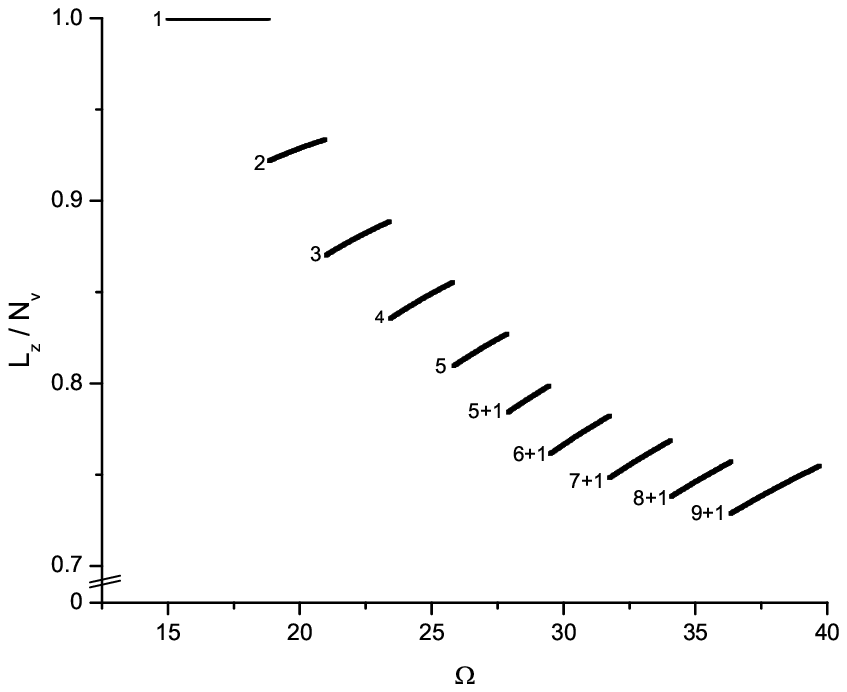}
\end{minipage}
\\
(a) uniform condensate
&
(b) TF condensate
\end{tabular}
\caption{Angular momentum per vortex 
as a function of the external rotation.  
Here, we assume $\ln\left(R/\xi\right)=8$.
}
\label{fig:LzPerNv}
\end{figure}

  It is interesting to consider the angular momentum per particle $L_z$,
which follows directly from the thermodynamic relation
\begin{equation}
L_z = -\left(\frac{\pt\mE'}{\pt\Omega}\right)_{\mathrm{eq}}.
\end{equation}
As a function of $\Omega$, 
the envelope of the minimum free energy curves 
is approximately proportional to $-\Omega^2$,
apart from small irregularities associated with the small number of vortices,
so that $L_z$ increases approximately linearly with $\Omega$.
For a given vortex state,
the angular momentum increases slowly with the external rotation
as the radius of the ring of vortices shrinks
to satisfy Eq.~(\ref{eq:r_r}).
Since the number of vortices also increases nearly linearly with $\Omega$,
it is natural to consider the ratio $L_z/N_v$, 
which is the angular momentum (per particle) per vortex.
This ratio $L_z/N_v$ is precisely 1 for a single central vortex.
For many vortices, in contrast, the velocity approaches 
$\mathbf{\Omega}\times\mathbf{r}$ (solid-body rotation).
In our dimensionless units, 
$L_z/N_v$ is then the average of $r^2$ 
weighted with the particle density $n(r)$,
which is $\frac{1}{2}$ for a uniform condensate 
and $\frac{1}{3}$ for the TF condensate
trapped in a harmonic potential.
Thus the ratio $L_z/N_v$ should start at 1 for both condensates 
and decrease slowly toward the appropriate limit with increasing $\Omega$.
This reduction of the angular momentum 
(per particle) per vortex from 1 was observed in \cite{Chev00}.

  Figure~\ref{fig:LzPerNv} shows our calculated 
angular momentum (per particle) per vortex ($L_z/N_v$)
as a function of the external rotation $\Omega$.
As expected from the above estimate,
the angular momentum per vortex of the TF condensate falls more rapidly
(typically by $\sim 1.4 - 1.6 \sim \frac{3}{2}$)
as the number of vortices increases.

\section{Tkachenko-like oscillation modes and the associated stability}
  Once we have an equilibrium ground state,
we can consider small excitations of the ground state
by considering small variations about the equilibrium.
The in-plane, nearly transverse excitation modes of the infinite vortex lattice
in the uniform Bose condensate are the well-known Tkachenko modes~\cite{Tkac66}.
Here we consider the finite-size analog of the Tkachenko modes
both in the bounded uniform condensate and in the trapped TF condensate.

  Let $\alpha_j$ and $\beta_j$ be the small variations
of the radial distance and the angle, respectively,
for the $j$th vortex in a ring,
\begin{equation}
\label{eq:variation}
r_j = r_r + \alpha_j, \qquad \phi_j = \phi_j^0 + \beta_j,
\end{equation}
where ($r_r, \phi_j^0$) is the equilibrium position of the $j$th vortex.
In equilibrium $N_r$ vortices are symmetrically placed in a ring:
$N_r = N_v$ for a state without a vortex at the center
and $N_r = N_v-1$ for a state with an extra vortex at the center;
the state has a $N_r$-fold rotational symmetry.
Thus a Fourier decomposition for $\bm{\alpha}$ and $\bm{\beta}$
will solve the problem:
\begin{equation}
\label{eq:Fourier}
\begin{pmatrix}
\alpha_j \\ \beta_j
\end{pmatrix}
   = \sum_{s=0}^{N_r-1}
         \begin{pmatrix}
         \tilde{\alpha}_s \\ \tilde{\beta}_s
         \end{pmatrix}
         e^{i2\pi js/N_r},
\end{equation}
where the Fourier index $s$ runs from 0 to $N_r-1$.

  The vortex at the center requires a special treatment.
Since polar coordinates become singular at the origin,
it is more convenient to use cartesian coordinates ($x$,$y$)
for the vortex at the center,
and polar coordinates ($r_j$,$\phi_j$)
for the rest of the vortices in the ring.
In this mixed-coordinate representation with $x$, $y$, $r_j$ and $\phi_j$,
the Euler-Lagrange equations of motion for the small variations become
\begin{equation}
\label{eq:motion}
\frac{d}{dt} \mathbf{d} = T_2^{-1} G E_{2} \mathbf{d},
\end{equation}
where $\mathbf{d}\equiv\left(x,y,\bm{\alpha},\bm{\beta}\right)$ is a vector.
The matrix $T_2=\mathrm{diag}\left(2,2,t,\cdots,t,t,\cdots,t\right)$,
with $t$ (it is independent of $q$) defined as
\begin{equation}
t \equiv \left(\frac{\pt^2\mT}{\pt r_q \pt \dot{\phi}_q}\right)_{\mathrm{eq.}},
\end{equation}
arises from the time-dependent part $\mT$ of the Lagrangian.
The matrix $E_2$ is the second-order expansion
of the free energy part $\Delta\mE'$
\begin{equation}
\left(E_2\right)_{ij}=
   \left(\frac{\pt^2\Delta\mE'}{\pt D_i \pt D_j}\right)_{\mathrm{eq.}},
\end{equation}
and here $\mathbf{D}\equiv\left(x,y,\mathbf{r},\bm{\phi}\right)$.
The antisymmetric matrix $G$ shows
how these small-variation parameters couple to
form the dynamical structure of the system:
\begin{equation}
G = \begin{pmatrix}
     0  & 1 & 0        & 0 \\
     -1 & 0 & 0        & 0 \\
     0  & 0 & 0        & I_{N_r} \\
     0  & 0 & -I_{N_r} & 0
     \end{pmatrix},
\end{equation}
where $I_{N_r}$ is the $N_r\times N_r$ identity matrix.
To apply the above equation~(\ref{eq:motion})
to the condensate with no central vortex,
ignore the first two components for $x$ and $y$ in $\mathbf{d}$
and set $x=y=0$ for the rest of the equations.

  Although these equations look complicated,
the Fourier decomposition~(\ref{eq:Fourier})
reduces them to a block-diagonal form,
because $\Delta\mE'$ includes only two-body terms
$\sim\int \mathbf{v}_i\cdot\mathbf{v}_j$
[see Eqs.~(\ref{eq:kinetic}), (\ref{eq:freeE_unif}) and (\ref{eq:freeE_TF})].
When the Fourier components are projected out,
the equations for $\tilde{\alpha}_s$ and $\tilde{\beta}_s$ 
simplify considerably.
For $N_r=N_v-1$ vortices in a ring and one at the center,
Eq.~(\ref{eq:Fourier}) and the orthogonality of the Fourier components
decouple the original $2N_v\times2N_v$ problem into
one $6\times6$ matrix problem,
\begin{equation}
\label{eq:6by6}
\frac{d}{dt}\mathbf{P}_{6} = M_{6}\,\mathbf{P}_{6},
\end{equation}
and $N_r-2$ sets of $2\times2$ matrix problems ($s=0,2,\ldots,N_r-2$),
\begin{equation}
\label{eq:2by2}
\frac{d}{dt}\mathbf{P}_{2}^s = M_{2}^s\,\mathbf{P}_{2}^s,
\end{equation}
where
$\mathbf{P}_{6}
   = (x,y,\tilde{\alpha}_1,\tilde{\alpha}_{N_r-1},
          \tilde{\beta}_1, \tilde{\beta}_{N_r-1})$
and $\mathbf{P}_{2}^s = (\tilde{\alpha}_s,\tilde{\beta}_s)$
are 6- and 2-component vectors, respectively.
If there is no vortex at the center,
the original problem corresponds to $N_r=N_v$ sets of
the above $2\times2$ matrix problems
[Eq.~(\ref{eq:2by2}) with $s=0,1,\ldots,N_r-1$].
The matrix elements of $M_{6}$ and $M_{2}^s$ are determined
from Eq.~(\ref{eq:motion}) (see Appendix~\ref{ap:matrix}).

  By solving these matrix problems,
we can study the stability of the vortex system.
For a given equilibrium state,
we determine the excitation spectrum
as a function of the external rotation frequency $\Omega$.
If a particular excitation frequency is real,
the corresponding small amplitudes
($x$,$y$,$\bm{\alpha}$,$\bm{\beta}$) oscillate around zero,
and the state is stable.
If any excitation frequency is complex with a positive imaginary part,
the corresponding mode will grow with time,
and the state is dynamically unstable.

\begin{figure}[h]
\centering
\begin{tabular}{cc}
\begin{minipage}{3in}
\includegraphics[height=2.4in,width=3in]{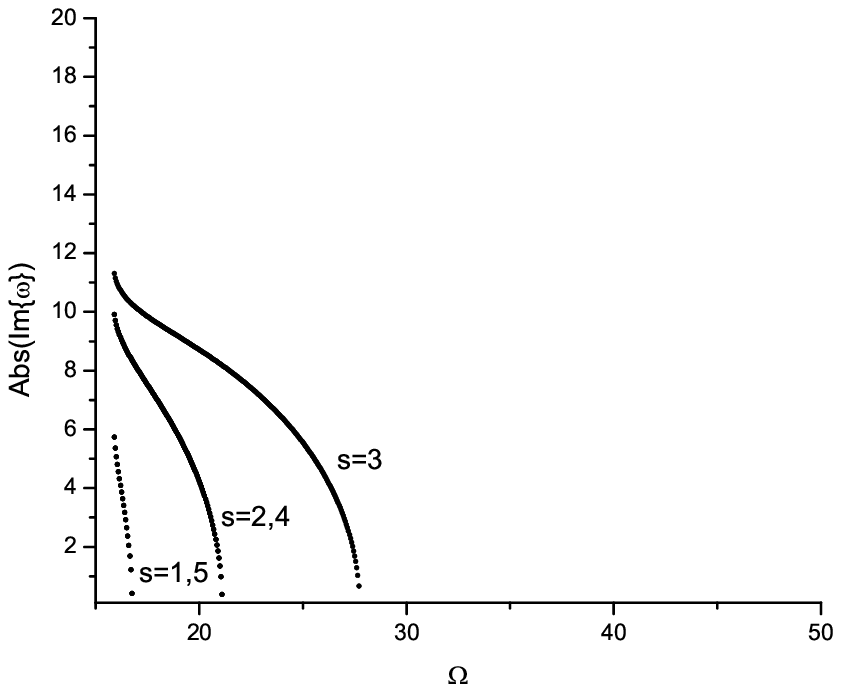}
\end{minipage}
&
\begin{minipage}{3in}
\includegraphics[height=2.4in,width=3in]{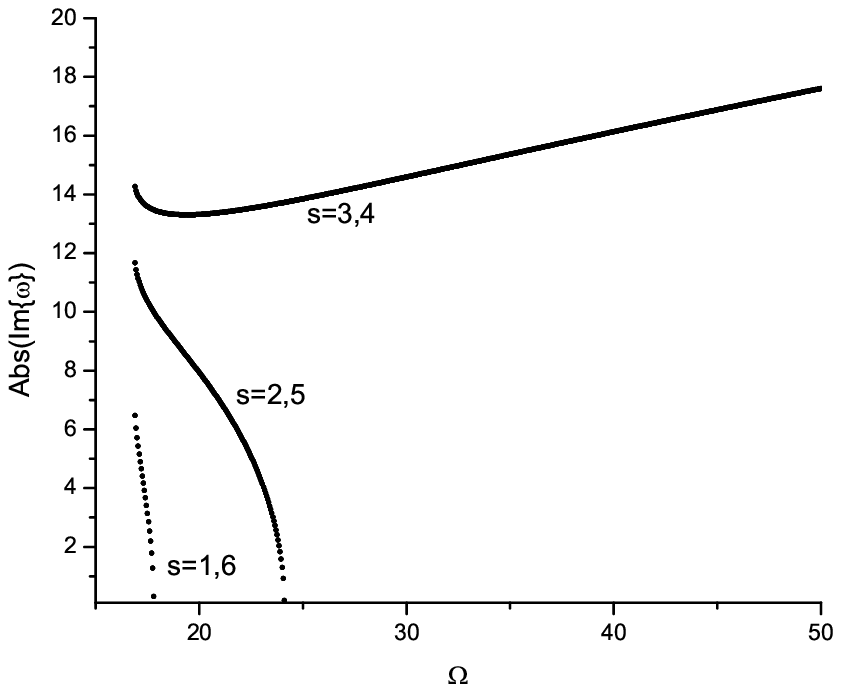}
\end{minipage}
\\
(a) six vortices in a ring
&
(b) seven vortices in a ring 
\end{tabular}
\caption{Imaginary parts of the excitation spectrum for a TF condensate
showing the normal-mode index $s$ for each curve.  
Here, we assume $\ln\left(R/\xi\right)=8$.
}
\label{fig:im_spectrum}
\end{figure}

  As an example, Fig.~\ref{fig:im_spectrum} shows
the positive imaginary parts of the excitation spectra
for the system of six and seven vortices
in a ring without a central vortex in the TF condensate.
Since the complex excitation frequencies always appear in conjugate pairs,
the figure shows only the positive imaginary parts.
Above a critical angular velocity ($\sim27.7$),
a ring of six vortices becomes stable.
In contrast, the ring of seven vortices is always unstable
because a mode with positive imaginary part persists for any $\Omega$.

  The Fourier index $s$ is analogous to the wave number in a linear array.
For a given $N_r$, 
the normal modes become increasingly unstable as $s$ increases,
so that the essential instability arises for short wave lengths.
In the case of six vortices,
$s=3$ (the mode with the shortest wave length) 
is the last mode to become stable.
In the seven-vortex state,
the shortest wave length modes are $s=(3~\text{or}~4)$,
and these modes remain unstable for all angular velocity, 
indicating a ``buckling'' transition to a `6+\tc{1}' configuration,
arising from the preference for a more uniform configuration.

\subsection{Uniform condensate}
  The dynamical instability of a finite number of vortices
in a single ring without a central vortex 
was studied by Havelock~\cite{Have31}.
There, by solving the set of $2\times2$ matrix problems~(\ref{eq:2by2}),
he concluded that a single ring is unstable 
when the number of vortices exceeds six.
Here, we sharpen this result.
Although six vortices in a ring are dynamically stable,
we find that the configuration with `5+\tc{1}'
has lower free energy compared to `6'.

\begin{figure}[h]
\centering
\begin{tabular}{cc}
\begin{minipage}{3in}
\includegraphics[height=2.4in,width=3in]{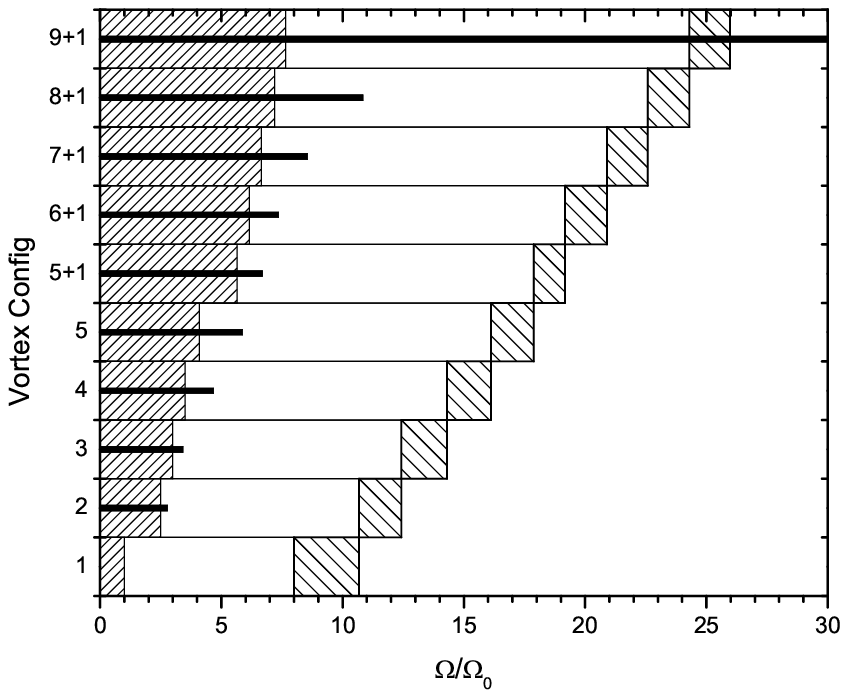}
\end{minipage}
&
\begin{minipage}{2.5in}
\includegraphics[height=2.4in,width=3in]{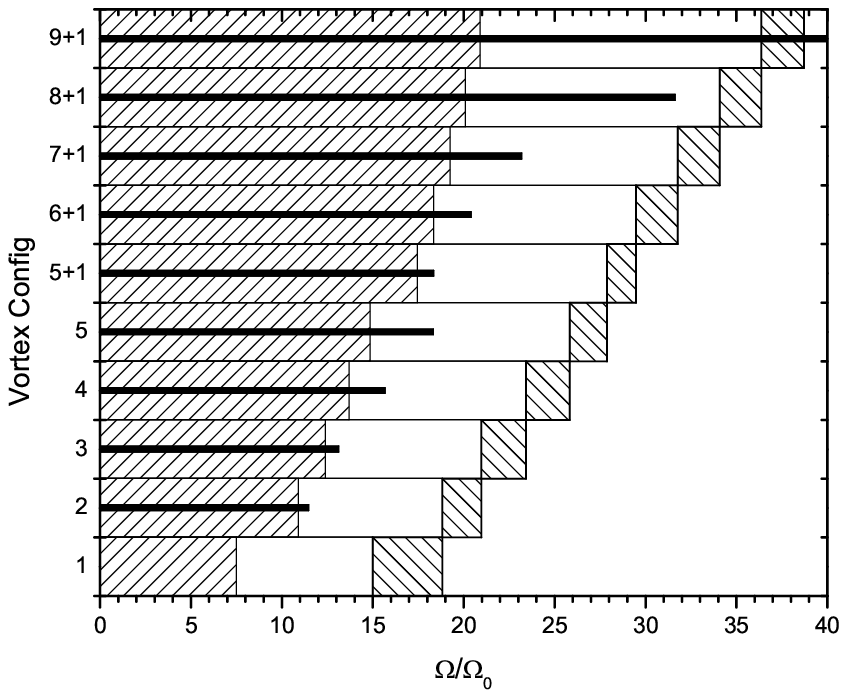}
\end{minipage}
\\
(a) uniform condensate
&
(b) TF condensate
\end{tabular}
\caption{Stability of single-ring states:
backslashed blocks denote the region with lowest free energy;
empty blocks denote metastable states that are local  free energy minima 
but energetically disfavored because other more stable states exist;
slashed blocks indicate regions where no local free energy minimum exists;
the narrow solid bar denotes the region with complex excitation  frequencies
(and thus dynamically unstable).
Here, we assume that $\ln\left(R/\xi\right)=8$.
}
\label{fig:vtx_window}
\end{figure}

  Figure~\ref{fig:vtx_window}~(a) summarizes
the conclusions of the free energy calculation
and the stability analysis in a uniform condensate in a rigid cylinder.
Here, the backslashed blocks show the regions
where a given configuration has the lowest free energy.
In the empty region to the left,
there exist energetically more stable states with fewer vortices.
For example, the empty region for the one-vortex state means
that the vortex-free state is preferred to the one-vortex state.
The slashed regions at the left side of the empty regions indicate
the absence of a local free energy minimum.
Hence, a state is metastable, but not globally stable,
in the empty region (to the right of the slashed region).
For each configuration, the narrow solid bar denotes the region
where the particular array has at least one complex excitation frequency
with a positive imaginary part (and is thus dynamically unstable).

  Unlike the ring with no central vortex,
the ring with a central vortex is stable
even for a ring of eight vortices.
The dynamical instability appears only
for nine vortices in the ring with a central vortex,
indicating the existence of other more stable states
than this `9+\tc{1}' configuration.
Campbell and Ziff~\cite{Camp79} studied
the stationary vortex patterns in a uniform condensate,
evaluating the corresponding free energy
for numbers of vortices $N_v=1$ to 30
(they did not investigate the small-amplitude normal modes).
According to their findings,
the stable state with the lowest free energy for a system of ten vortices,
has `4+4+2' vortices
(three concentric rings; beginning from the outer-most ring);
the `9+\tc{1}' configuration has slightly higher free energy
and is nearly stable, as opposed to stable.

\subsection{TF condensate}
  The TF condensate shows qualitatively the same behavior
as the uniform condensate case, with small differences
[see Fig.~\ref{fig:vtx_window}(b)].
 
\begin{enumerate}
\item   For a single vortex, the dynamics is dominated 
        either by the image vortex (for the uniform condensate)
        or by the nonuniform trap potential (for the TF condensate).
        For two or more vortices, however, 
        the vortex-vortex interactions become increasingly dominant, 
        and the two systems are expected to become more and more similar.
\item   In a transition from `5' to `5+\tc{1}' for the TF condensate,
        the region of complex excitation modes (indicated by the solid bar)
        remains almost unchanged,
        although the number of vortices increases.
\item   Another notable difference is that,
        for the `8+\tc{1}' TF configuration,
        the region of complex eigenmodes (dynamically unstable)
        expands significantly, 
        leaving a narrower region of external rotation frequency
        over which this configuration is metastable.
\end{enumerate}

\section{discussion}
  The time-dependent variational analysis based on the Lagrangian formalism
provides a direct and flexible method
to study the dynamics of vortices in a Bose-Einstein condensate.
By analyzing the time-independent free energy $\mE'$,
we determine the various ground states
and the associated angular velocities
for the transitions from one state to the next.
A more complete dynamical picture emerges
when we consider the full Lagrangian $\mL = \mT - \mE'$,
for it provides the explicit time dependence:
for example, the vortex precession frequency,
the excitation spectrum and the dynamical stability.

  For a given configuration of vortices,
the critical $\Omega$ for the onset of stability 
is larger for the TF condensate than for the uniform condensate,
by approximately a factor of two.
As a result, a single ring of $N_r$ vortices (without and with a central vortex)
has a smaller radius in the TF condensate than in the uniform condensate,
typically, by $\sim 0.9$.

  The Lagrangian formalism has the notable feature
that it apparently incorporates valuable physical information
that would otherwise emerge only with significant additional effort.
For example, the Lagrangian analysis of the low-lying normal modes
of a condensate~\cite{Pere96} used a Gaussian trial function,
but the resulting dynamics reproduced the known frequencies
even in the TF limit where the density profile has a parabolic form.

  Similarly here, the Lagrangian formalism yields
the correct precession of a single vortex,
even though the assumed stream function
is equivalent to purely incompressible flow.
To understand this question in more detail,
note that the velocity field assumed in Eq.~(\ref{eq:stream})
can equivalently be written as
$\mathbf{v} = - \bm{\nabla}\times (\hat{\mathbf{z}}\chi)$,
which implies that ${\bm{\nabla}\cdot\mathbf{v}}= 0$.
Since we are merely constructing a trial function,
this choice is acceptable,
even though the correct equation of continuity requires
${\bm{\nabla} \cdot (n\mathbf{v})}=0$.
Nevertheless, the Lagrangian approach incorporates
sufficient physical information (presumably through the energy $\mE'$)
that the resulting precession in Eq.~(\ref{eq:prec_TF})
agrees with other more lengthy analyses.

  A direct approach to the precession relies on the full equation
${\bm{\nabla} \cdot (n\mathbf v)}
   = {\mathbf v}\cdot \bm \nabla n
     + n{\bm \nabla \cdot \mathbf v}= 0$.
The extra term involving $\bm{\nabla} n$
(namely the spatial variation in the condensate density)
produces additional logarithmic contributions
to the phase~\cite{Svid00,Svid00a,Shee04},
which drive the precessional motion.
Thus the precession of a single vortex in a TF condensate
can be considered to arise from the nonuniform trap potential
and the resulting nonuniform TF density.
It would be interesting to improve the trial function
by including these additional logarithmic contributions explicitly.
Since the variational Lagrangian approach is ``exact''
for the exact solution of the GP equation,
such an improved wave function should provide an improved description,
but it is not clear whether the dynamical precession rate would be affected.

  As studied in this work, the single-ring configuration
is applicable only for a small number of vortices
($N_v\lesssim 9$ in practice)
in a TF condensate that rotates relatively slowly.
The single ring does have one other very interesting application, however,
when the condensate has a central hole and thus becomes annular.
Such behavior is predicted to occur in a rapidly rotating trap
with an additional quartic confinement,
which allows the angular velocity $\Omega$
to exceed the harmonic trap frequency $\omega_\perp$.
When the width of the annulus becomes sufficiently narrow,
the condensate is predicted to have a one-dimensional array of vortices
arranged in a single ring~\cite{Kasa02,Fett04}.
The techniques developed here should apply directly
to the dynamics and stability of this interesting configuration.

\begin{acknowledgments}
  ALF is grateful to the Kavli Institute for Theoretical Physics,
University of California, Santa Barbara,
where part of this work was completed.
The research at KITP was supported in part
by the National Science Foundation under Grant No. PHY99-0794.
\end{acknowledgments}

\appendix
\section{evaluating the TF free energy difference}
\label{ap:TFfree}
  With no loss of generality, consider the case with two vortices at
$\mathbf{r}_1=(r_1,\phi_1)$ and $\mathbf{r}_2=(r_2,\phi_2)$.
The stream function that describes the velocity field in the condensate is
\begin{equation}
\chi=\chi_1+\chi_2
     =\ln\left|\mathbf{r}-\mathbf{r}_1\right|
      + \ln\left|\mathbf{r}-\mathbf{r}_2\right|.
\end{equation}
In the free energy difference $\Delta\mE'=\Delta\mE-\Omega L_z$,
the angular momentum is straightforward to evaluate
and only $\Delta\mE$ is of concern.
\begin{equation}
\begin{split}
\Delta\mE
   &= \frac{1}{2}\int d^2\mathbf{r}~ n(r) \mathbf{v}^2 \\
   &= \frac{1}{2}\int d^2\mathbf{r}~ n(r)
              \left(\left|\bm\nabla\chi_1\right|^2
                    + \left|\bm\nabla\chi_2\right|^2
                    + 2 \bm\nabla\chi_1\cdot\bm\nabla\chi_2\right).
\end{split}
\end{equation}

  The first two terms are the contributions of each vortex,
and the last term is the interaction between the two vortices.
For the first term, an integration by parts gives
\begin{equation}
\begin{split}
\frac{1}{2}\int d^2\mathbf{r}~ n \left|\bm\nabla\chi_1\right|^2
   &= \frac{1}{2}\int d^2\mathbf{r}~ 
            \bm\nabla\cdot\left(n\chi_1\bm\nabla\chi_1\right)
      - \frac{1}{2}\int d^2\mathbf{r}~ 
            \bm\nabla n\cdot\left(\chi_1\bm\nabla\chi_1\right)
      - \frac{1}{2}\int d^2\mathbf{r}~ 
            n\chi_1\nabla^2\chi_1.
\end{split}
\end{equation}

  Excluding the healing length $\xi$ neighborhood around the vortex core,
the last term in the integral vanishes
because $\mathbf{r}_1$ lies outside the domain of the integral
[$\nabla^2\chi_1=2\pi\delta^{(2)}(\mathbf{r}-\mathbf{r}_1)$].
Correspondingly, the surface integral (the first term) has two boundaries,
a TF radius and a $\xi$ neighborhood.
With the expansion of the stream function
in angular harmonics (\ref{eq:harmonics}),
the integral becomes
\begin{equation}
\frac{1}{2}\int d^2\mathbf{r}~ n \left|\bm\nabla\chi_1\right|^2
   = \frac{1}{2}\left(1-r_1^2\right)
     \left[2\ln\left(\frac{R}{\xi}\right)+\ln\left(1-r_1^2\right)\right]
      + \frac{1}{2}\left(2r_1^2-1\right).
\end{equation}
A similar approach can be used to evaluate the interaction term.

\section{matrix elements of the equations of motion}
\label{ap:matrix}
  The orthogonality of the Fourier components simplifies
the matrix elements of the equations of motion.
\begin{equation}
M_{6} =
\begin{pmatrix}
\frac{1}{2}\Delta\mE'_{,yx}   & \frac{1}{2}\Delta\mE'_{,yy} &
-i\frac{1}{2} N_r E           & i\frac{1}{2} N_r E&
\frac{1}{2} N_r F             & \frac{1}{2} N_r F
\\
-\frac{1}{2}\Delta\mE'_{,xx}  & -\frac{1}{2}\Delta\mE'_{,xy} &
-\frac{1}{2} N_r E            & -\frac{1}{2} N_r E &
-i\frac{1}{2} N_r F           & i\frac{1}{2} N_r F
\\
it^{-1} F                     & t^{-1} F&
C_1                           & 0 &
D_1                           & 0
\\
-it^{-1} F                    & t^{-1} F &
0                             & C_{N_r-1} &
0                             & D_{N_r-1}
\\
t^{-1} E                      & -it^{-1} E &
-A_1                          & 0 &
-B_1                          & 0
\\
t^{-1}E                       & it^{-1} E &
0                             & -A_{N_r-1} &
0                             & -B_{N_r-1}
\end{pmatrix},
\end{equation}
and
\begin{equation}
M_{2}^s = \begin{pmatrix}
             C_s &  D_s \\
            -A_s & -B_s
          \end{pmatrix},
\end{equation}
where
\begin{align}
A_s &= \frac{1}{t}\sum_{j=1}^{N_r}\Delta\mE'_{,r_q r_j}
                                  e^{i(\phi_j-\phi_q) s}, \\
B_s &= \frac{1}{t}\sum_{j=1}^{N_r}\Delta\mE'_{,r_q \phi_j}
                                  e^{i(\phi_j-\phi_q) s}, \\
C_s &= \frac{1}{t}\sum_{j=1}^{N_r}\Delta\mE'_{,\phi_q r_j}
                                  e^{i(\phi_j-\phi_q) s}, \\
D_s &= \frac{1}{t}\sum_{j=1}^{N_r}\Delta\mE'_{,\phi_q \phi_j}
                                  e^{i(\phi_j-\phi_q) s},
\end{align}
and $\Delta\mE'_{,pq}$ means
\begin{equation}
\Delta\mE'_{,pq} 
  \equiv \left(\frac{\pt^2\Delta\mE'}
                    {\pt p\,\pt q}\right)_{\mathrm{eq.}}.
\end{equation}
For a uniform condensate,
$E=1+r^{-2}$,
$F=-1+r^{-2}$,
$t=2r$;
for a TF condensate,
$E=-\left(1-2r^{-2}+\ln r^2\right)$,
$F=-r\left(1-2r^{-2}-\ln r^2\right)$,
$t=2r\left(1-r^2\right)$.


\end{document}